\documentstyle{article}
\begin{document}
\begin{center}

{\LARGE \bf 
One loop renormalization of the 4-D Quantum Dilaton Gravity
in 
Spherically symmetric background }

\vspace{15mm}

{\sc Seiji Mukaigawa}
\footnote{E-MAIL: mukaigaw@theo.phys.sci.hiroshima-u.ac.jp}
\\
Department of Physics, Hiroshima University,
Higashi-Hiroshima 739, Japan. \\

\vspace{1cm}

{\sc Hiroyuki Takata}
\footnote{E-MAIL: takata@theory.kek.jp ,
          URL: http://theory.kek.jp/~takata/}\\
National Laboratory for High Energy Physics (KEK),
Tsukuba, Ibaraki 305, Japan.
and 
Department of Physics, Hiroshima University,
Higashi-Hiroshima 739, Japan. \\

\vspace{5.5cm}

{\bf Abstruct}
\end{center}
\baselineskip 8mm

We study the one loop renormalization in the most general
metric-dilaton theory with second derivative only.
In constant background dilaton theory, there are two types of
gravity background which enable the theory renormalizable at
one-loop level. We show this concretely to discuss on the
spherical symmetric background.

\pagebreak
%%%%%%%%%%%%%%%%%%%%%%%%%%%%% document %%%%%%%%%%%%%%%%%%%%%%%%%%%%%%%%%%

\def\beq{\begin{equation}}
\def\eeq{\end{equation}}
\def\bce{\begin{center}}
\def\ece{\end{center}}
\def\ben{\begin{enumerate}}
\def\een{\end{enumerate}}
\def\ul{\underline}
\def\ni{\noindent}
\def\nn{\nonumber}
\def\bs{\bigskip}
\def\ms{\medskip}
\def\wt{\widetilde}
\def\wh{\widehat}
\def\Tr{\mbox{Tr}\ }

%\maketitle

%%%%%%%%%%%%%%%%%%%%%%%%%%%%% section 1 %%%%%%%%%%%%%%%%%%%%%%%%%%%%%%%%%
The construction of quantum theory of gravity is one of the
fundamental problems in the theoretical physics.
The experimental check of Einstein gravity provides good
result, but this theory is not renormalizable.
In the absence of matter fields and cosmological constant
case Einstein's gravity is the finite theory at one loop
level. %\cite{k2} 
That is, the effective action is finite on
mass shell at one loop level. 
But this theory diverges in two loop order. %\cite{k3,k4}
The higher derivative gravity with the terms quadratic in
curvature tensor is renormalizable % \cite{s15}
 and allow the
renormalization group study of some physical effect like
asymptotic freedom % \cite{s16,s17,s18}
 and phase transitions, % \cite{s28,s29} 
 but not unitary because the ghosts and
tachyons are present in the spectrum of the theory.
%\cite{k7-k11}
Furthermore the pure Einstein gravity with a cosmological
constant is renormalizable,  % \cite{s20}
 however one introduces
matter fields the one loop renormalizability is lacking 
even on mass shell.
On the other hand in Supergravity theory, due to the
presence of local supersymmetry, the theory is two loop
finite, but the divergent terms are present in the three
loop order % \cite{k14,k15}

Recently it has been a considerable interest to the
metric-scalar gravity in four dimensions.
The active research in this field was inspired by string
theory.
The effective action of string (or
superstring) depends on both metric and dilaton.
That effective action arise in a form of the power series in
a slope parameter $\alpha'$, and the point of view is that
the higher orders in such expansion correspond to higher
energies. 
From this point of view at lower energy scale the action for
gravity has the form of the lower derivative dilaton action. 
Probably the completely consistent quantum theory of gravity
can be constructed within the string theory, and gravity
will be described by effective action within this frame.
However the string theory can be valid at the Planckian
energy and above, and if one would like to deal with the
energy below that scale, it is natural to suppose that the
quantum effects of gravity will be related with some low
energy action.
Thus at the moment we do not have any consistent theory that
is applicable below that energy scale and any research
in this field is based on the choice of some model, which
allow us to  explore some gravity effects.

In this paper we consider the four dimensional metric-dilaton
model with second derivative terms only. 
The action is
following :  
\beq
S= \int d^4x \sqrt{-g}\; \{ A(\phi)g^{\mu\nu}\partial_{\mu}\phi
\partial_{\nu}\phi + B(\phi)R -2B(\phi)\Lambda(\phi) \}
\label{action}
\eeq
where $A(\phi),B(\phi)$ and $\Lambda(\phi)$ are arbitrary
functions of $\phi$. 

If the consideration is restricted by one loop on shell
case, then the theory with cosmological term $\Lambda(\phi)$ can
be renormalizable that leads to some general conjectures
about the high energy behavior of quantum gravity
\cite{s16}.
Next we can restrict ourselves by some special backgrounds
where the theory is renormalizable.
On the other hand one can introduce an additional constraint
on the background dilaton and regard it as constant.
This way is also of some cosmological interest , because of
the renormalizability in the potential sector enables one to
evaluate the significance of quantum gravity for the
cosmological phase transitions.
The theory (1) has already researched on the constant
background dilaton theory in the arbitrary gravitational
background case \cite{shaptaka} and maximal symmetric space-time
background case \cite{taka}.
In this paper we describe one in the spherically symmetric
space-time background case and show simultaneously that
there are two types of background that enable this theory
renormalizable.
%%%%%%%%%%%%%%%%%%%%%%%%%%% section 2 %%%%%%%%%%%%%%%%%%%%%%%%%%%%%%%%%

For the purpose of calculation of the divergences we will
apply the background field method and the Schwinger-DeWitt technique.
Let us start with the usual splitting of the fields
into background $g_{\mu\nu}, \phi$
and quantum $h_{\mu\nu}, \varphi$ ones
\beq
\phi \rightarrow \phi' = \phi + \varphi\;,\;\;\;\;\;\;\;\;\;\;\;\;\;\;
\;   g_{\mu\nu} \rightarrow g'_{\mu\nu} + h_{\mu\nu}         \label{2.8}
\eeq
The one-loop
effective action is given by the standard general expression
\beq
\Gamma={i \over 2}\;\Tr\ln{\hat{H}}-i\;\Tr\ln {\hat{H}_{ghost}},\label{2.9}
\eeq
where $\hat{H}$ is the bilinear form of the action (1) with
added gauge fixing term and $\hat{H}_{ghost}$
is the bilinear form of the gauge ghosts action.
To perform the calculations in a most simple way one needs to
introduce the special form of the gauge fixing term:
\beq
S_{gf} = \int d^4 x \sqrt{-g}\;\chi_{\mu}\;\frac{\alpha}{2}\;\chi^{\mu}
                                                           \label{2.10}
\eeq
where $\chi_{\mu} = \nabla_{\alpha} \bar{h}_{\mu}^{\,\alpha}+
\beta\nabla_{\mu}h+\gamma \nabla_{\mu} \varphi$, $h=h_{\mu}^{\mu},\;
\bar{h}_{\mu\nu}=h_{\mu\nu}-\frac{1}{4}\;hg_{\mu\nu}$ and $\alpha, \beta,
\gamma$ are some functions of the background dilaton, which can be tuned
for our purposes. For instance, if one choose these functions as
follows
\beq
\alpha=-B\;\;,\;\;\;\;\beta=-\frac{1}{4}\;\;,\;\;\;\;\gamma=-\frac{B_1}{B}
                                                        \label{2.11}
\eeq
then the bilinear part of the action $S+S_{gf}$ and the operator $\hat{H}$
has especially simple (minimal) structure
$$
\left(S + S_{gf}\right)^{(2)}
=\int d^4 x \sqrt{-g}\; {\omega} \hat{H} {\omega}^T
$$
\beq
\hat{H}=\hat{K}\Box+\hat{L}_{\rho}\nabla^{\rho}+\hat{M}    \label{2.12}
\eeq
Here $\omega=\left(\bar{h}_{\mu\nu},\;h,\; \varphi\right)$ and $T$ means
transposition. The components of $\hat{H}$ have the form
\[
\hat{K}=\left(
\begin{array}{ccc}
              \frac{B}{4} \delta^{\mu \nu \alpha \beta} & 0 & 0\\
              0 & -\frac{B}{16} & -\frac{B_1}{4} \\
              0 & -\frac{B_1}{4} & \frac{B_1^2}{2B} -A
\end{array}
\right)
\]
\[
\hat{L}^{\lambda}=\left(
\begin{array}{ccc}
     \begin{array}{c}
          \frac{B_1}{4} \{ \delta^{\mu \nu \alpha \beta}
          g^{\tau \lambda} \\
          +2 g^{\nu \beta} (  g^{\mu \tau } g^{\alpha \lambda }
          - g^{\alpha \tau } g^{\mu \lambda } ) \}
     \end{array}
     & - \frac{B_1}{4} g^{\mu \tau} g^{\nu \lambda}
     & \left( \frac{B_2}{2}-A \right) g^{\mu \tau} g^{\nu \lambda}\\
     \frac{B_1}{4} g^{\alpha \tau} g^{\beta \lambda}
     & -\frac{B_1}{16} g^{\tau \lambda}
     & \left(\frac{A}{4} -\frac{5}{8} B_2 \right) g^{\tau \lambda}\\
      \left( A - \frac{B_2}{2}\right) g^{\alpha \tau} g^{\beta \lambda}
     & \left( \frac{B_2}{8}-\frac{A}{4}\right) g^{\tau \lambda}
     & \left(\frac{B_1^2}{2B} -A\right)_1 g^{\tau \lambda}
\end{array}
\right) (\nabla_{\tau}\phi)
\]
\[
\hat{M}=
\left( 
\begin{array}{ccc}
   \begin{array}{l}
        \delta^{\mu \nu \alpha \beta}
        \{ \frac{B_1}{2}(\Box \phi) \\
        +( \frac{B_2}{2}-\frac{A}{4} )(\nabla \phi)^2
         +\frac{B\Lambda}{2} \} \\ 
         + g^{\nu \beta} \{
         -B_1 ( \nabla^{\mu} \nabla^\alpha \phi ) \\
         + ( A-B_2 )(\nabla^\mu \phi)(\nabla^\alpha \phi ) \}\\
         +\frac{B}{4} \{ -\delta^{\mu \nu \alpha \beta} R \\
         +2 g^{\nu \beta} R^{\mu \alpha}
         +2R^{\mu \alpha \nu \beta} \}
    \end{array}
    \!\!\!\! 
    & \!\!\!\! 0 \!\!\!\! 
    & \!\!\!\! 
    \begin{array}{l}
       \frac{B_2}{2} ( \nabla^{\mu} \nabla^{\nu} \phi ) \\
        + \left( \frac{B_3}{2} - \frac{A_1}{2} \right)
        (\nabla^\mu \phi)(\nabla^\nu \phi) \\
        - \frac{B_1}{2}R^{\mu \nu}
    \end{array} \\
    \!\!\!\! & \!\!\!\!
    \!\!\!\! & \!\!\!\! \\
    \begin{array}{l}
        \frac{B_1}{4} ( \nabla^{\alpha} \nabla^{\beta} \phi ) \\
        +\frac{B_2}{4} (\nabla^\alpha \phi)(\nabla^\beta \phi)
    \end{array}      
    & \!\! -\frac{B\Lambda}{8}
    & \!\! 
    \begin{array}{l}
        -\frac{3}{8} B_2 (\Box \phi) \\
        + ( \frac{A_1}{8}-\frac{3}{8}B_3 )(\nabla \phi)^2 \\
        + \frac{B_1}{8} R - \frac{(B\Lambda)_1}{2}
    \end{array} \\
    \!\!\!\! 
    & \!\!\!\!
    \!\!\!\! 
    & \!\!\!\! \\
    \begin{array}{l}
        A ( \nabla^{\alpha} \nabla^{\beta} \phi ) \\
        +\frac{A_1}{2} (\nabla^\alpha \phi)(\nabla^\beta \phi) \\
        - \frac{B_1}{2}R^{\alpha \beta}
    \end{array}    
    &\begin{array}{l}
         -\frac{A}{4} (\Box \phi) \\
         +\frac{A_1}{8} (\nabla \phi)^2 \\
         +\frac{B_1}{8} R-\frac{(B\Lambda)_1}{2}
     \end{array}
    &\begin{array}{l}
         -A_1(\Box \phi)\\
         -\frac{A_2}{2}(\nabla \phi)^2 \\
         +\frac{B_2}{2} R - (B\Lambda)_2
     \end{array}
\end{array}
\right)
\]
The next problem is to separate the divergent part of $\Tr\ln\hat{H}$.
To make this we rewrite this trace in a following way.
\beq
\Tr \ln\hat{H}  =\Tr\ln\hat{K}+
\Tr\ln\left(\hat{1}\Box + \hat{K}^{-1} \hat{L}^{\mu}\nabla_\mu
+\hat{K}^{-1}\hat{M} \right)                            \label{2.15}
\eeq
One can notice that the first term does not give contribution to
the divergences. Let us explore  the second term which has standard
minimal form and can be easily estimated with the use of standard
Schwinger-DeWitt method \cite{DW,hove}. 

The bilinear form of the ghost action also has the minimal structure
\beq
\hat{H}_{ghost}=
g^{\mu \alpha}\Box+\gamma(\nabla^{\alpha}\phi)\nabla^{\mu}
+ \gamma (\nabla^{\mu} \nabla^{\alpha} \phi) + R^{\mu \alpha} \label{2.16}
\eeq
and it's contribution to the divergences can be easily derived with the use of
the standard methods.

The most general off-shell structure of the one-loop
divergence of the effective action has the form :
\beq
\begin{array}{ccl}
\Gamma_{div}^{1-loop} & =-\frac{1}{16 \pi^2 (n-4)} \int d^4x\sqrt{-g}[ &
 c_e G + c_w C^2 + c_r R^2 + c_4 R(\nabla \phi)^2 + c_5
R(\Box \phi ) \\
&& + c_6 R^{\mu\nu}(\nabla_{\mu} \phi)
(\nabla_\nu \phi) + c_7 R + c_8 (\nabla \phi)^4  \\
&& + c_9 (\nabla \phi)^2(\Box \phi) + c_{10} (\Box \phi)^2 
+ c_{11}(\nabla \phi)^2 + c_{12} ] \\ 
& + (s.t.) & \\
\end{array}
\eeq
where $C^2=C_{\mu\nu\alpha\beta }C^{\mu\nu\alpha\beta}$ is the square of Weyl
tensor and$(\nabla\phi)^2=g^{\mu\nu}\;\nabla_\mu\phi\;\nabla_\nu\phi$. $"s.t."$
means ``surface terms''. Expression for the coefficient
functions $c$ are in the reference \cite{shaptaka}.

Let us now make some comments concerning the above result. The one loop
divergences (9) essentially depend on the choice of the
functions $A(\phi), B(\phi), \Lambda(\phi)$ in the starting action.
In particular, this dependence concerns the $c_w$ and $c_r$ functions,
which correspond to the terms with the second powers in curvature tensor.
The above expressions are valid only in the case
$\;X=2A\,B\,-3B_{1}^{2}\neq 0$.
For $X=0$ the calculational scheme must be modified because of extra
conformal symmetry \cite{shaptakac}.

We want to restrict to coefficient function $A,B$ and $\Lambda$ to
take the background which follows the classical equations of
motion.
The equation of motion for $g_{\mu\nu}$ and $\phi$ are
following :
\beq
\left\{
\begin{array}{c}
R_{\mu\nu} - \frac{1}{2}R g_{\mu\nu} + \Lambda g_{\mu\nu} = T_{\mu\nu} \\
B_{1}R - 2(B\Lambda)_{1} - A_{1}(\nabla\phi)^{2} -
2A(\Box\phi) = 0
\end{array}
\right.
\eeq
where 
\beq
T_{\mu\nu} = - 
\left\{ (\frac{B_{2}}{B}-\frac{A}{2B})(\nabla\phi)^{2} +
\frac{B_{1}}{B}(\Box\phi) \right\}g_{\mu\nu} + 
(\frac{B_{2}}{B} - \frac{A}{B})(\nabla_{\mu})(\nabla_{\nu})
+ \frac{B_{1}}{B}(\nabla_{\mu}\nabla_{\nu}\phi).
\eeq
It is difficult to solve these equations, so we put $\phi$
constant and impose spherical symmetry for gravity.\\

{\bf \underline{ step 1 }}  Putting $\phi = const.$ \\

To simplify this problem we make the equation (10) to have the
solution $\phi = constant$.
Then we see 
\beq
T_{\mu\nu} = 0.
\eeq
Substituting this equation into Einstein equation (10) and
contracting the equation we obtain 
\beq
R = 4 \Lambda .
\eeq
Next Using the equation of motion for $\phi$ we obtain
\beq
B = c \Lambda .
\eeq
where $c$ is a arbitrary constant. 
Thus $B$ is expressed in $\Lambda$ and vise versa.
One the other hand since $c_{e}$ becomes constant, 
$\sqrt{-g}c_{e}G$ is total derivative term and it can be
ignored.Hence the one-loop divergent (9) becomes following
expression 
\beq
\Gamma^{1-loop}_{div,\phi=const.} =
-\frac{1}{16\pi^{2}(n-4)}\int d^{4}x \sqrt{-g}[
 c_{w}C_{\mu\nu\rho\sigma}^{2} +
\frac{371}{90 c^{2}}B(\phi)^{2} ].
\eeq

{\bf \underline{ step 2 }}  Imposing the spherical symmetric condition \\

The metric can be written in following form when the
4-dimensional space-time has 3-dimensional isometry groups
and its orbit makes 2-dimensional surface :
\beq
ds^{2} = -\epsilon e^{2\nu}(dx^{0})^{2} +
e^{2\lambda}(dx^{1})^{2}
+ r^{2}(d\theta^{2} + \epsilon f^{2}(\theta)d\phi^{2}).
\eeq
If the orbit is space-like(time-like) then $\epsilon =
+1(-1)$.
$f(x)$ has following function corresponding to the sign of 
sectional curvature $K$ of the orbit
\beq
f(x) = \left\{
\begin{array}{cc}
\sin x  & (K>0) \\
   x    & (K=0) \\
\sinh x & (K<0) \\ 
\end{array}
\right..
\eeq
Choosing $x_{0},x_{1}$ suitable 
we can set $\lambda = -\nu$ without any loss in generality.
\beq
ds^{2} = -e^{2\nu}dt^{2} + e^{-2\nu}dr^{2} +
r^{2}d\sigma_{2}^{2}
\eeq
where $d\sigma_{2}^{2}$ is the metric of the 2-dimensional
constant curvature space with sectional curvature 
$k=0,\pm1$ ($\epsilon=1$).

We give the expression for $e^{2\nu}$ in equation (18).
The well-known solution of the Einstein equation with
$\Lambda \neq 0 , T_{\mu\nu} = 0$ in spherical symmetry
is Schwarzschild-deSitter space.
the solution is following :
\beq
e^{2\nu} = 1-\frac{k}{r}-\frac{1}{3}\Lambda r^{2}
\eeq
where $k$ is a constant.
If we introduce electromagnetic field as external field,
there is additive term from generalized Birkhoff's theorem
\beq
\epsilon \frac{k'}{r^{2}}
\eeq
where $k'$ is a constant.
It is easy to check
\beq
C^{\mu\nu\rho\sigma}C_{\mu\nu\rho\sigma} \neq 0. 
\eeq
Thus this term makes nonrenormalizable structure.
So we choose $c_{w} =0$. This equation becomes following :
\beq
86AB-249B_{1}^{2} = 0.
\eeq
Thus $A$ is expressed in $B$ and vice versa.

From above considerations, we find that the one loop
divergence of effective action on the spherically 
symmetric background is same as the maximally symmetric
case.
The expression has following form : 
\beq
\Gamma^{1-loop}_{div,Spherical} = 
\frac{1}{16\pi^{2}\epsilon}\int d^{4}x \sqrt{-g}
\left[ -\frac{371}{90 c^{2}} B(\phi)^{2} \right].
\eeq
Difference between maximally symmetric case and spherically
symmetric case of the structure of divergence is that 
function $A$ and $B$ are not independent.
that is ,the free function is only one in three input
functions $A$ ,$B$ and $\Lambda$ .
And we can perform normalization in the same way as in
the maximally symmetric case.
bare function $B_{0}$ is related in the following
relation :
\beq
B_{0}(\phi) = \mu^{\frac{\epsilon}{2}}
(1+\frac{1}{16\pi^{2}\epsilon}\frac{371}{360
c})B(\phi)
\eeq
where $\mu$ is the renormalization scale.

In constant dilaton background theory, there are two types of
gravitational background. the number of free
function is determined whether the gravitational background
has non-zero $C_{\mu\nu\rho\sigma}^{2}$ or not.
If $C_{\mu\nu\rho\sigma}^{2} = 0$ then the number of free
function is 2. And if $C_{\mu\nu\rho\sigma}^{2} \neq 0$ then
the number of free function is 1.
In the maximal symmetric case, the number of free function
is 2, and in spherically symmetric case, the number is 1.
In arbitrary background case, there is no free functions.
that is,the form of $A,B$ and $C$ are determined except for
it's parameter.
   
%%%%%%%%%%%%%%%%%%%%%%%%% bibliography %%%%%%%%%%%%%%%%%%%%%%%%%%%%%


\begin{thebibliography}{99}
%%%%%%%%%%%%%%%%%%%%%%%%%%%%%%%%%%%%%%%%%%%%%%%%%%%%%%%%%%%%%
% Some macros are available for the bibliography:
%   o for general use
%      \JL : general journals          \andvol : Vol (Year) Page
%   o for individual journal 
%      \PR  : Phys. Rev.               \PRL : Phys. Rev. Lett.
%      \NP  : Nucl. Phys.              \PL  : Phys. Lett.
%      \JMP : J. Math. Phys.           \CMP : Commun. Math. Phys.
%      \PTP : Prog. Theor. Phys.       \JPSJ: J. Phys. Soc. Jpn.
%      \JP  : J. of Phys.              \NC  : Nouvo Cim.
%      \IJMP: Int. J. Mod. Phys.       \ANN : Ann. of Phys.
% Usage:
%   \PR{D45,1990,345}            ==> Phys.~Rev.\ {\bf D45} (1990), 345
%   \JL{Phys.~Lett.,A30,1981,56} ==> Phys.~Lett.\ {\bf A30} (1981), 56
%   \andvol{B123,1995,1020}      ==> {\bf B123} (1995), 1020
%%%%%%%%%%%%%%%%%%%%%%%%%%%%%%%%%%%%%%%%%%%%%%%%%%%%%%%%%%%%%
%\bibitem{GSW} M. B. Green, J. H. Schwarz and E. Witten, 
%{\sl Superstring Theory}, 
%(Cambridge University Press, Cambridge, 1987). 

%\bibitem{St} K. S. Stelle, 
%{\sl Renormalization of the Higher Derivative Quantum Gravity}, 
%Phys. Rev. {\bf 16D}, (1977), 953. 

\bibitem{s16} E.S.Fradkin and A.A.Tyutin,
Nucl.Phys.{\bf 201B},469(1982).

\bibitem{shaptaka} I. L. Shapiro and H. Takata, 
{\sl One Loop Renormalization of the Four-Dimensional Theory for Quantum Dilaton Gravity}, 
Phys. Rev.{\bf D52}, 2162, (1995).

\bibitem{taka} H.Takata, hep-th/9604170.
{\sl 4-D Quantum Dilaton Gravity During Inflation and
Renormalization at one loop}

\bibitem{DW} B. S. DeWitt, 
{\sl Dynamical Theory of Groups and Fields}, 
(Gordon and Breach, NY, 1965).

\bibitem{hove}G. t'Hooft and M. Veltman, 
{\sl One Loop Divergences in the Theory of Gravitation}, 
Ann. Inst. H. Poincare. {\bf A20}, 69, (1974).

\bibitem{shaptakac} I. L. Shapiro and H. Takata, 
{\sl Conformal Transformation in Gravity}, 
Phys. Lett. {\bf B361}, (1995), 31.

\bibitem{BOS} I. L. Buchbinder, S. D. Odintsov and I. L. Shapiro, 
{\sl Effective Action in Quantum Gravity}, 
(IOP, Bristol, 1992).

\end{thebibliography}
\end{document}